\documentclass[iop]{emulateapj}
\begin{document}

\title{THE SPECTROSCOPIC ORBITS OF FIVE SOLAR TYPE, SINGLE LINED BINARIES}


\author{FRANCIS C. FEKEL,\altaffilmark{1} SAMIRA RAJABI, MATTHEW W. 
MUTERSPAUGH, AND MICHAEL H. WILLIAMSON}
\affil{Center of Excellence in Information Systems, 
    Tennessee State University, 3500 John A. Merritt Boulevard,
    Box 9501, Nashville, TN 37209}
\email{fekel@evans.tsuniv.edu, samira@coe.tsuniv.edu, matthew1@coe.tsuniv.edu,
michael.h.williamson@gmail.com}

\altaffiltext{1}{Visiting Astronomer, Kitt Peak National Observatory,
National Optical Astronomy Observatory, operated by the Association
of Universities for Research in Astronomy, Inc. under cooperative agreement
with the National Science Foundation.}

\begin{abstract}
We have determined spectroscopic orbits for five single-lined 
spectroscopic binaries, HD 100167, HD 135991, HD 140667, HD 158222,
HD 217924. Their periods range from 60.6 to 2403 days and 
the eccentricities, from 0.20 to 0.84. Our spectral classes for the 
stars confirm that they are of solar type, F9 to G5, and all are dwarfs.  
Their [Fe/H] abundances, determined spectroscopically, are close to 
the solar value and on average are 0.12 greater than abundances from 
a photometric calibration.  Four of the five stars are rotating 
faster than their predicted pseudosynchronous rotational velocities.
\end{abstract}


\section{INTRODUCTION}
Nearly a decade ago a group of solar-type stars with $V$ magnitudes
between 5.9 and 8.1 was selected to expand a photometric 
program that was designed to detect long-term spot cycles in dwarfs 
similar to the Sun \citep{h99}.  Although most of these moderately 
bright stars had radial velocities published in the literature, many 
did not.  To remove binaries from the photometric sample, we 
obtained spectra and measured radial velocities.  Our spectroscopic 
observations soon identified five single-lined binaries, HD~100167, 
HD~135991, HD~140667, HD~158222, and HD~217924, which we continued 
to observe.  The basic properties of these binaries are given in 
Table~\ref{tbl-basic}.  During the course of our extended 
spectroscopic observing campaign, others have also identified 
these five stars as binaries.  From our observations we determine 
spectroscopic orbits and spectral and luminosity classes, measure 
iron abundances and $v$~sin~$i$ values, and then briefly discuss the 
systems. 
  
\section{OBSERVATIONS AND RADIAL VELOCITIES}
From 2002 through 2012 we obtained observations at the Kitt Peak 
National Observatory (KPNO) with the coud\'e feed telescope and coud\'e 
spectrograph.  The vast majority were acquired with a TI~CCD detector, 
and those spectrograms were centered at 6430~\AA, cover a wavelength 
range of 84~\AA, and have a resolution of 0.21~\AA\ or a resolving power 
of just over 30,000.  The TI CCD was unavailable in 2008 September, and 
so a Tektronics CCD, designated T1KA, was used instead.  With that CCD 
the spectrum was centered at 6400~\AA.  Although the wavelength range 
covered by the chip increased to 172~\AA, the resolving power decreased 
to 19,000.  Beginning in 2010 September we acquired spectra with a CCD,
consisting of a 2600$\times$4000 array of 12 $\mu$m pixels, that was 
made by Semiconductor Technology Associates and designated STA2.
With STA2 the spectrum was once again centered at 6430~\AA, and the much 
larger size of the detector produced a wavelength range of 336~\AA. The
spectrograph slit was set so that the STA2 spectra have the same
resolution as those acquired with the TI CCD although there is
some worsening of the resolution at both ends of the STA2 spectra.  All
our KPNO spectra have signal-to-noise (S/N) ratios of about 150.

Starting in 2006, we collected additional spectrograms with the 
Tennessee State University (TSU) 2 m automatic spectroscopic telescope 
(AST) and a fiber-fed echelle spectrograph, situated at Fairborn
Observatory near Washington Camp in the Patagonia Mountains of 
southeastern Arizona \citep{ew04,ew07}.  Through 2011 June the detector 
was a 2048$\times$4096 SITe ST-002A CCD with 15~$\mu$m pixels. 
\citet{ew07} have discussed reduction of the raw spectra and their 
wavelength calibration.  The AST echelle spectrograms have 21 orders 
that cover the wavelength range 4920--7100~\AA, and most have an average 
resolution of 0.17~\AA\ although a few have a slightly lower resolution 
of 0.24~\AA.  The S/N ratios ranged from 30 to 80, being affected in part 
by significant temperature fluctuations of the dewar. 

Beginning in 2010 January, we made several upgrades to the TSU 2 m
AST to increase throughput, sensitivity, and flexibility.  First, an 
entirely new focal plane assembly and guide system were constructed 
and installed, including a replacement for the fiber optic links 
between the telescope's focal plane and spectrographs.  The new 
system increases the number of available fibers from three to a
maximum of 12, each terminated with industry-standard SMA connectors, 
and enables the switching between fibers to occur in a matter of 
seconds. This configuration expands our flexibility and allows us to 
host guest instruments.  The fibers connect to custom 1-inch aluminum 
mirrors, and the connector tine is flush with the front mirror 
surface.  A nonpolarizing cubic customized 70/20 beamsplitter is 
placed before the mirror/fiber tip to redirect the smaller portion 
of starlight to a small CCD camera for direct guiding; starlight 
reflected back from the mirror/fiber connector is redirected by the 
beamsplitter in the opposite direction, where a second CCD camera 
can similarly be used as a ``slit monitor'' and for guiding on 
brighter stars.  Second, an aluminum secondary mirror replaced the 
telescope's existing metal mirror, because the latter's protective 
coating had very significantly eroded with time.  Third, in 
the summer of 2011 the SITe CCD detector and dewar were replaced with 
a Fairchild 486 CCD having 4K$\times$4K 15~$\mu$m pixels, which 
required a new readout electronics package, and a new dewar with a
Cryotiger refrigeration system, 
which is consistently cooled at $-110 \, {\rm ^\circ C}$.  Overall, 
these improvements have increased the throughput by at least a factor of  
30.  The echelle spectrograms that were obtained with this 
new detector have 48 orders, covering the wavelength range 3800--8260~\AA.  
Because different diameter fibers were used at times, the resolution was 
either 0.24 or 0.4~\AA, which resulted in S/N ratios 
that ranged from 60 to 130 at 6000~\AA.  Finally, a new temperature
controlled building has been constructed to host guest instruments.  It 
has two rooms, one, which measures 6 feet by 30 feet, to house the 
electronics systems and a second, which is 11 feet by 30 feet, to 
accommodate optics systems.

The KPNO velocities were determined by cross-correlation with respect to 
IAU radial velocity standard stars of similar spectral type, $\beta$~Vir,
HR~7560, and $\iota$~Psc.  The velocities adopted for those standards 
are from \citet{s10}.

\citet{ftw09} described the measurement of the radial velocities
from the Fairborn Observatory AST spectra.  Unlike the KPNO velocities 
those from Fairborn Observatory are absolute velocities.  Our 
unpublished velocities of several IAU solar-type standard stars 
indicate that the Fairborn Observatory velocities taken with the SITe 
CCD have a small zero-point offset of $-$0.3 km~s$^{-1}$ relative to 
the velocities of \citet{s10}.  Starting in the fall of 2011, 
velocities from spectra obtained with the new CCD system have a 
zero-point offset $-$0.6 km~s$^{-1}$ relative to those of \citet{s10}.
Thus, we added either 0.3 or 0.6 km~s$^{-1}$, depending on which 
detector was used, to each measured velocity so that the zero point 
of the Fairborn velocities is the same as that of the KPNO velocities.

\section{GENERAL ORBITAL ANALYSIS}
For each of the five stars we first found a preliminary orbital period 
with the program PDFND, which uses the least string method, implemented 
by T. J. Deeming \citep{betal70}, to identify the period.  We then 
computed a preliminary orbit for each star with the program BISP 
\citep{wetal67}, which uses a slightly modified version of the 
Wilsing-Russell method.  We refined those orbits with SB1 
\citep{betal67}, a program that computes differential corrections.  

For each star we determined separate solutions of the KPNO velocities 
and the Fairborn Observatory velocities.  From the ratio of the variances 
of the solutions, we then determined weights for the velocities from the 
two observatories and produced a final orbital solution, listed in 
Table~\ref{tbl-orbele}, with the combined set of velocities.

\section{SPECTRAL TYPE}
\citet{sf90} identified several luminosity-sensitive and temperature-sensitive
line ratios in the 6430-6465~\AA\ region.  They employed those critical line
ratios and the general appearance of the spectrum as spectral-type criteria.  
However, for stars that are hotter than about early-G spectral class, the
line ratios in that wavelength region have little sensitivity to luminosity.
In those cases we have used the entire 84~\AA\ spectral region of our KPNO 
observations to estimate just the spectral classes of the individual 
components.  The luminosity class may be determined by computing the 
absolute visual magnitude with the {\it Hipparcos} parallax and 
comparing that magnitude to evolutionary tracks or a table of canonical 
values for giants and dwarfs.  Such a luminosity class that is not 
derived from the actual spectrum we note as `dwarf' rather than with 
the Roman numeral luminosity class. Our classifications are summarized 
in Table~\ref{tbl-spty}.

\section{PROJECTED ROTATIONAL VELOCITY}
We have determined $v$~sin~$i$ values from our red-wavelength
KPNO spectra with the procedure of \citet{f97}.  Following \citet{f97}, for 
late-F and G stars, we adopted a macroturbulent broadening value of 
3 km~s$^{-1}$.  The resulting projected rotational velocities are 
listed in Table~\ref{tbl-spty}.  Those $v$~sin~$i$ values have estimated
uncertainties of 1 km~s$^{-1}$. 

\section{ABUNDANCE ANALYSIS}
We determined the iron abundance for each of the five stars from our
Fairborn Observatory echelle specta.  Our line list 
was taken from \citet{bubar10} and includes between 34 and 45 Fe~I lines
for each star. All abundances were derived from equivalent width (EW)
measurements with the use of the IRAF {\texttt {splot}} package.
Table~\ref{tbl-EW} shows the line list for each star along with the
excitation potential (EP), oscillation strength (log$gf$), and measured
width for each line.
We obtained stellar model atmospheres for the stars by
interpolating non-overshoot models in the ATLAS9 grid \citep{ck03}.
The initial effective temperature $T_{eff}$ and log$gf$ values are 
from \citet{casa11}. We adopted the solar abundance of  
iron as the starting value in the atmospheric model for each
star and chose an initial microturbulent velocity, $v_{t}$, of 1.5 
km~s$^{-1}$. We used the model atmosphere as input to the LTE line
abundance analysis code MOOG, developed by \cite{sned73}. We then
refined the parameters iteratively until there was no trend in Fe I
abundance as a function of reduced equivalent width or as a function
of EP. An increase of 100 K in the adopted temperature increases the
abundance by 0.06 dex. The best fit model parameters for each star 
and the derived 
[Fe/H] abundances and standard deviation ($\sigma$) values are 
listed in Table~\ref{tab::atmosabund}.

\subsection{Results comparison}
All five of our binaries are included in the Geneva-Copenhagen survey 
of solar neighborhood stars \citep{netal04}, which provided basic
properties, including abundance estimates.
\citet{hetal07} derived an improved photometric iron abundance 
calibration using Str\"omgren photometry and redetermined the iron 
abundances for the stars in that survey.  In Table~\ref{tbl-abundcomp} 
we compare their adopted effective temperatures and revised
[Fe/H] values with our results. On average our effective
temperatures are 155 $\pm$ 18 K higher resulting in the [Fe/H] values
being 0.12 $\pm$ 0.03 greater than those of \citet{hetal07}.
One of our binaries, HD~100167, has a previously determined 
spectroscopic iron abundance.  \citet{vf05} found $T_{eff}$ = 5915 K 
and [Fe/H] = 0.06 for that star, values which are essentially 
identical to ours.

\section{CIRCULARIZATION AND SYNCHRONIZATION}
The two main theories of orbital circularization and rotational
synchronization (e.g., \citealp{z77,tt92}) disagree significantly
on absolute time scales but do agree that synchronization should occur
first.  
Observationally, \citet{dm91} examined the multiplicity of solar-type 
stars in the solar neighborhood.  They determined that while systems 
with periods $\leq$10 days had circular orbits, longer period orbits are 
generally eccentric. With periods ranging from 60.6 to 2403 days it is
not surprising that the five binaries have significant eccentricities,
which range from 0.20 for HD~140667 to 0.845 for HD~217924. 

Obviously, in an eccentric orbit true synchronization can not occur.  
However, for a non-circular orbit \citet{h81} has shown that the rotational 
angular velocity of a star will tend to synchronize with that of the 
orbital motion at periastron, a condition called pseudosynchronization.  
With equation (42) of \citet{h81}, we compute pseudosynchronous periods
of 10.4, 41.8, 781.6, 97.0, and 132.2 days for HD 100167, HD 135991, 
HD 140667, HD 158222, and HD 217924, respectively.

\section{INDIVIDUAL STARS}
\subsection{HD 100167 = HIP 56257}
\subsubsection{Brief history}
\citet{netal02} included HD 100167 [$\alpha$ = 11$^h$ 31$^m$ 53.91$^s$
$\delta$ =  41\arcdeg 26\arcmin 21.8\arcsec (2000)] in a velocity survey 
of late-type stars.  From two observations they found it to have a 
variable velocity with an average of $-$27.6 km~s$^{-1}$. \citet{netal04} 
listed HD 100167 in another extensive compilation, the 
Geneva-Copenhagen survey of solar neighborhood stars. From two Coravel 
observations they determined an average velocity of $-$28.1 km~s$^{-1}$ 
and also identified the star as a binary.  HD~100167 was initially 
considered for the Keck planet search program, and from a single 
spectrogram \citet{vf05} found a velocity of $-$29.5~km~s$^{-1}$.
\citet{wetal07} observed HD~100167 twice as part of yet another survey.
Their work supported the $Spitzer$ Legacy Science Program ``Formation and
Evolution of Planetary Systems.''  Velocities of their two spectra differ
by 3.4 km~s$^{-1}$ but have an average velocity of $-$27.3 km~s$^{-1}$, 
similar to previous results.

Both \citet{netal04} and \citet{vf05} determined the metallicity and
rotational velocity of HD~100167.  \citet{netal04} found [Fe/H] = $-$0.05,
revised by \citet{hetal07} to $-$0.09, and $v$~sin~$i$ = 5 km~s$^{-1}$, 
while \citet{vf05} got [Fe/H] = 0.06 and $v$~sin~$i$ = 3.8 km~s$^{-1}$.

Based on a comparison with a grid of stellar evolutionary tracks, 
\citet{hetal09} determined an age of 6.8 Gyr for HD 100167, while 
\citet{tetal07} found a likely age of 3.4 Gyr.  \citet{wetal04} obtained
Ca II H and K emission measurements from high resolution spectra and 
determined an age of 2.3 Gyr. From the more recent activity--age 
calibration of \citet{mh08} (their equation (3)) the age is slightly 
increased to 2.7 Gyr.  Also using stellar activity indicators, 
\citet{cetal09} assigned a younger age of 1.58 Gyr for HD~100167.
 
\subsubsection{Orbital analysis}
With 142 Fairborn Observatory velocities all phases of the very 
eccentric orbit of HD~100167 are well sampled.  However, our 23 KPNO 
radial velocities only partly cover the orbit with much of the very 
eccentric portion of the velocity curve missing.  Thus, in computing
an orbit for the KPNO velocities we fixed the semiamplitude at the
value from the Fairborn Observatory solution.  The center-of-mass
velocities of the Fairborn and KPNO solutions differ by just 0.1 
km~s$^{-1}$, and the other elements are also in very good agreement, 
so we combined the two data sets, assigning weights of 1.0 and 0.15 
to the Fairborn and KPNO velocities, respectively, and computed a
final orbital solution.  Table~\ref{tbl-100167} provides a list of 
our velocities, and the final orbital elements from the joint data 
solution are given in Table~\ref{tbl-orbele}.  Figure~1 compares 
the velocities with the radial velocity curve computed from the 
final orbital elements.

\subsubsection{Discussion}
From a comparison with various reference stars we assign a spectral 
class of G0 to HD~100167, while the $Hipparcos$ parallax and apparent
magnitude (Table~1) combine to indicate that the star is a dwarf.  
The mass function, computed 
from the orbital elements, is 0.049 $M_{\sun}$, more than twice as
large as that for any of the other four binaries.  Such a value 
is large enough to suggest that lines of the secondary might be visible 
at red wavelengths \citep[e.g.,][]{sf92}.  However, upon 
careful examination of our Fairborn and KPNO spectra, no lines of a 
secondary were detected, indicating that the magnitude difference 
between the components is greater than 2.5 \citep{sf92}.

From a calibration of Ca~II H and K emission measurements \citet{wetal04} 
obtained a rotation period of 16 days compared with our pseudosynchronous
period of 10.4 days.  With the basic parameters in Table~\ref{tbl-basic} 
we obtain a radius of 1.08~$R_{\sun}$ from the Stefan-Boltzmann relation. 
Then the 16 day period results in a rotational velocity of 3.4 km~s$^{-1}$, 
while the shorter pseudosynchronous period produces a larger velocity of 
5.3 km~s$^{-1}$.  The measured $v$~sin~$i$ value of 4 km~s$^{-1}$, 
averaged from our result and two other independent determinations 
\citep{netal04,vf05}, is intermediate between the two predictions.  

\subsection{HD 135991 = HIP 74821}
\subsubsection{Brief history}
\citet{netal04} also included HD 135991 [$\alpha$ = 15$^h$ 17$^m$ 27.62$^s$ 
$\delta$ = 22\arcdeg 17\arcmin 59.5\arcsec \\ (2000)]
in their Geneva-Copenhagen survey 
of solar neighborhood stars.  From four spectrograms they determined its
radial velocity to be $-$22.1 km~s$^{-1}$.  \citet{retal07} observed
the star as part of an extensive low-resolution spectroscopic survey for 
metal-rich planet-search targets.  From several calibrated spectral 
indices they found HD~135991 to have [Fe/H] = $-$0.13, while the photometric
calibration of \citet{hetal07} produced [Fe/H] = $-$0.20. 

\subsubsection{Orbital analysis }
Orbital phase coverage of both the 95 Fairborn and 34 KPNO radial velocities
of HD~135991 is very good. Our separate solutions of the two data sets 
produced excellent agreement between all orbital elements, including the two 
center-of-mass velocities, which agreed to better than 0.1 km~s$^{-1}$.  
We next computed a final joint solution in which all velocities were 
given unit weight.  Our velocities are listed in Table~\ref{tbl-135991}, 
and the final orbital elements from the joint data solution are given in 
Table~\ref{tbl-orbele}.  Figure~2 provides a comparison of our 
velocities with the radial velocity curve computed from the final 
orbital elements.

\subsubsection{Discussion}
Our spectral class is F9 and the $Hipparcos$ parallax and apparent magnitude
from Table~1 combine to identify the star as a dwarf.  We determine a 
$v$~sin~$i$ value of 6.7 km~s$^{-1}$ 
in reasonable agreement with the value of 8 km~s$^{-1}$ found by 
\citet{netal04}.  Of our five binaries HD~135991 has the largest
projected rotational velocity and the second shortest pseudosynchronous 
rotation period, 41.8 days. From the Stefan--Boltzmann relation we 
adopt a radius of 1.17~R$_{\sun}$.  The period and radius result in a 
pseudosynchronous rotational velocity of 1.4 km~s$^{-1}$, while HD~135991 
is actually rotating at least five times more rapidly. 
 
\subsection{HD 140667 = HIP 77098}
\subsubsection{Brief history}
HD 140667 [$\alpha$ = 15$^h$ 44$^m$ 31.86$^s$ $\delta$ =  11\arcdeg 
15\arcmin 59.4\arcsec (2000)] was classified by \citet{ht70} as a 
G0~V star.  \citet{o94} obtained Str\"omgren photometry of it, and 
\citet{netal04} included it in the Geneva-Copenhagen survey of stars 
in the solar neighborhood.  From three observations their average 
radial velocity is $-$16.7 km~s$^{-1}$.

\subsubsection{Orbital analysis}
We acquired 123 Fairborn and 25 KPNO radial velocities of HD 140667, and 
both sets of observations provide good coverage of the orbital phases.
The orbital elements from individual solutions of the two data sets are 
in very good agreement, with the two center-of-mass velocities 
differing by less than 0.1 km~s$^{-1}$.  All observations were given 
unit weights in a final solution of the combined data.  
Table~\ref{tbl-140667} lists our velocity observations, and the orbital 
elements from our combined solution are given in Table~\ref{tbl-orbele}.  
The radial velocities and the radial velocity curve from our orbital 
elements are shown in Figure~3.

\subsubsection{Discussion}
From our red wavelength spectra we determine a spectral class of F9 for 
HD~140667. Its {\it Hipparcos} parallax and apparent magnitude 
(Table~1) lead us to the conclusion that the star is a dwarf. Our 
classification is in close agreement with the
G0~V spectral type of \citep{ht70}.  We measure a $v$~sin~$i$ value of
2.3 km~s$^{-1}$ compared with a value of 4 km~s$^{-1}$ by \citet{netal04}.
HD~140667 has the longest pseudosynchronous rotation period of our five 
binaries, 782 days, resulting in a predicted pseudosynchronous rotational 
velocity of 
less than 0.1 km~s$^{-1}$, and so this star, although rotating slowly,
is clearly rotating faster then its pseudosynchronous rate.

\subsection{HD 158222 = HIP 85244}
\subsubsection{Brief history}
Like HD~100167, \citet{netal02} observed HD 158222 [$\alpha$ = 17$^h$ 25$^m$ 
08.32$^s$ $\delta$ =  53\arcdeg 07\arcmin 57.9\arcsec (2000)] in a velocity 
survey of late-type stars.  From three observations they found it to have a 
variable velocity with an average of $-$14.3 km~s$^{-1}$.  This value 
is very different from the average value of $-$27.5 km~$^{-1}$ found 
by \citet{netal04}. 

Based on a comparison with a grid of stellar evolutionary tracks,
\citet{hetal09} determined an age of 10.2 Gyr for HD 158222.
\citet{wetal04} obtained Ca II H and K emission measurements from high 
resolution spectra and used a chromospheric activity calibration to
determine a much younger age of 4.1 Gyr, while the more recent activity--age
calibration of \citet{mh08} (their equation (3)) increases the age  
to 4.9 Gyr.

\subsubsection{Orbital analysis}
With 152 Fairborn Observatory radial velocities all phases of the moderately
eccentric orbit of HD~158222 are well sampled.  Although the 25 KPNO 
velocities do not cover a small phase region around periastron passage, they 
still produce a good orbit.  Indeed, the orbital elements of the Fairborn 
and KPNO solutions are in excellent agreement, and in particular, the 
center-of-mass velocities differ by less than 0.1 km~s$^{-1}$.  
With weights of 1.0 for the Fairborn velocities and 0.7 for the KPNO 
velocities, a solution of the combined data sets was obtained.
Table~\ref{tbl-158222} lists our velocity observations, and the final
orbital elements are given in Table~\ref{tbl-orbele}.  Figure~4 
compares the radial velocities and the radial velocity curve from our 
orbital elements.

\subsubsection{Discussion}
Comparing the spectrum of HD~158222 with those of various reference stars 
of known spectral type, we classify it as G5~V.  Our $v$~sin~$i$ value
of 2.2 km~s$^{-1}$ is in excellent agreement with the value of
2 km~s$^{-1}$ from \citet{netal04}. The pseudosynchronous rotation period
for HD~158222 is 97 days, and from the Stefan--Boltzmann relation 
we estimate a radius of 1.09~$R_{\sun}$. Those values result in a 
pseudosynchronous rotational velocity of 0.6 km~s$^{-1}$, and so
HD~158222, although slowly rotating, is rotating faster than that value.

\subsection{HD 217924 = HIP 113884}
\subsubsection{Brief history}
\citet{ht70} classified HD~217924 [$\alpha$ = 23$^h$ 03$^m$ 50.59$^s$
$\delta$ = 21\arcdeg 35\arcmin 53.8\arcsec (2000)] as G0~V.
Like the previous four binaries \citet{netal04} included HD~217924 
in their extensive survey of solar neighborhood stars.
From three Coravel observations they found an average velocity of 
19.1 km~s$^{-1}$ but with a standard deviation of 3.8 km~s$^{-1}$. They 
concluded from the velocity variations that HD~217924 is a spectroscopic 
binary.  They also estimated [Fe/H] = $-$0.34 and $v$~sin~$i$ = 
3~km~s$^{-1}$.   As a result of a comparison of the {\it Hipparcos}
and Tycho-2 proper motions, \citet{fetal07} also flagged it as a binary. 

\subsubsection{Orbital analysis}
Given the very large eccentricity of its orbit, the phase coverage, 
and the small number of orbital cycles that have been observed, there 
are two possible periods for HD~217924, 2648 and 2403 days, depending 
on whether the first KPNO velocity is placed on the descending or 
ascending part of the velocity curve, respectively.  With the longer 
period all the other KPNO velocities that were obtained before the 
Fairborn observations were begun have large systematic residuals from 
the orbital fit. Comparing the velocity residuals for orbits  
with the above two periods, the sum of the weighted square 
of the residuals decreases by 45\% for the orbit with the shorter 
period. Thus, we have identified the period of 2403 days or 
6.6 yr as the correct one. This period is by far the longest of the 
five binaries that we have analyzed. 

Our 109 Fairborn velocities nearly cover one orbital period and are well 
distributed in phase except for the very rapid nodal passage, which 
mostly took place when HD 217924 was behind the Sun.  Fortunately, two 
observations were obtained after the star began its rapid velocity change 
from maximim to minimum velocity, significantly constraining the orbital 
elements.  The 17 KPNO velocities extend over nearly 1.5 orbital cycles 
and provide moderately good orbital phase coverage, but unfortunately,
they completely miss the maximum velocity peak and the rapid velocity 
decline through periastron passage.  As a result, those velocities by 
themselves do not enable the orbital solution of the elements to converge.
However, adopting the elements from the Fairborn velocities results 
in an excellent orbital solution of the KPNO velocities.  The two
sets of velocities are listed in Table~\ref{tbl-217924}.
With weights of 1.0 and 0.3 for the Fairborn and KPNO velocities, 
respectively, these data produce the orbital solution given in 
Table~\ref{tbl-orbele}.  Our radial velocities and the computed
radial velocity curve are compared in Figure~5.
 
\subsubsection{Discussion}
Our spectral type for HD~217924 is G2~V, in reasonable accord with
the G0~V type of \citet{ht70}.  Our projected rotational velocity
of 2.2~km~s$^{-1}$ is similarly consistent with the value of 
3~km~s$^{-1}$ determined by \citet{netal04}.  
Despite its 6.6 yr orbital
period, the orbital eccentricity of HD~217924 is so large, 0.84, 
that the pseudosynchronous rotation period is just 132 days.
Nevertheless, such a period is long enough that like all but one of
the other solar-type stars in this work, HD~217924 is rotating faster
than its pseudosynchronous velocity.

\acknowledgments
Astronomy research at Tennessee State University is supported in part
by NSF grant AST-1039522 and state of Tennessee through its Centers 
of Excellence program. This research made use of the SIMBAD database, 
operated at CDS, Strasbourg, France.



\clearpage

\begin{deluxetable}{lccccc}
\tablenum{1}
\tablewidth{0pt}
\tablecaption{Basic Properties \label{tbl-basic}}
\tablehead{
\colhead{} & \colhead{} & \colhead{$V$\tablenotemark{a}}
& \colhead{$B-V$\tablenotemark{a}} & \colhead{Parallax\tablenotemark{b}}  
 \\ 
\colhead{HD} & \colhead{HIP}  &
\colhead{(mag)} & \colhead{(mag)}
& \colhead{(mas)} 
}
\startdata
100167 & 56257  & 7.35 & 0.617  &  28.26 $\pm$ 0.72   \\ 
135991 & 74821  & 7.84 & 0.578  &  19.56 $\pm$ 0.91   \\
140667 & 77098  & 7.53 & 0.611  &  30.39 $\pm$ 1.54   \\
158222 & 85244  & 7.82 & 0.667  &  24.05 $\pm$ 0.65   \\
217924 & 113884 & 7.22 & 0.631  &  38.37 $\pm$ 2.11   \\
\enddata
\tablecomments{}
\tablenotetext{a}{\citet{petal97}}
\tablenotetext{b}{\citet{vl07}}
\end{deluxetable}



\begin{deluxetable}{lcccccccc}
\tabletypesize{\footnotesize}
\tablenum{2}
\tablewidth{0pt}
\tablecaption{Orbital Elements and Related Parameters 
\label{tbl-orbele}}
\tablehead{\colhead{HD} & \colhead{$P$} & \colhead{$T$} & \colhead{$e$}
& \colhead{$\omega$} & \colhead{$K$} & \colhead{$\gamma$} & \colhead{$a$~sin~$i$}
& \colhead{$f(m)$} \\
\colhead{} & \colhead{(days)} & \colhead{(HJD)} & \colhead{} & \colhead{(deg)}
& \colhead{(km~s$^{-1}$)} & \colhead{(km~s$^{-1}$)} & \colhead{(10$^6$ km)}
& \colhead{($M_{\sun}$)}
}
\startdata
100167 & 60.58037 & 2454239.1125 & 0.67827 & 20.661 & 26.975
& $-$20.218 & 16.512 & 0.04900 \\
  & $\pm$0.00030 & $\pm$0.0067 & $\pm$0.00047 & $\pm$0.072 & $\pm$0.029
& $\pm$0.013 & $\pm$0.020 & $\pm$0.00018  \\
135991 & 151.0054 & 2454262.42 & 0.5697 & 135.44 & 8.057 & $-$24.916 & 13.749
& 0.004542 \\
  & $\pm$0.0094 & $\pm$0.11 & $\pm$0.0022 & $\pm$0.40 & $\pm$0.027 & $\pm$0.020
& $\pm$0.053 & $\pm$0.000052  \\
140667 &978.42 & 2454060.0 & 0.2040 & 202.1 & 4.725 &$-$16.727
& 62.23 & 0.01003 \\
  & $\pm$0.94 & $\pm$3.2 & $\pm$0.0049 & $\pm$1.2 & $\pm$0.022 & $\pm$0.015
& $\pm$0.30 & $\pm$0.00014  \\
158222 &206.088 & 2454309.15 & 0.4126 & 78.94  & 10.994 &$-$25.549 & 28.380
& 0.02145 \\
  & $\pm$0.018  & $\pm$0.16 & $\pm$0.0017 & $\pm$0.30 & $\pm$0.022 & $\pm$0.015
& $\pm$0.062 & $\pm$0.00014  \\
217924 & 2402.7 & 2455181.29 & 0.8449   & 58.16 & 6.778 & 16.572 & 119.8 & 0.01187      \\
  & $\pm$17.7 & $\pm$0.67 & $\pm$0.0036 & $\pm$0.53 & $\pm$0.057 & $\pm$0.017 &  $\pm$1.9 & $\pm$0.00049  \\
\enddata
\end{deluxetable}


\begin{deluxetable}{lcc}
\tablenum{3}
\tablewidth{0pc}
\tablecaption{Spectral Type and $v$~sin~$i$ \label{tbl-spty}}
\tablehead{ \colhead{HD} & \colhead{Spectral} & \colhead{$v$~sin~$i$} \\
\colhead{} & \colhead{Type\tablenotemark{a}} & \colhead{(km~s$^{-1}$)}
}
\startdata
 100167   & G0 dwarf & 3.2 \\
 135991   & F9 dwarf & 6.7 \\
 140667   & F9 dwarf & 2.3 \\
 158222   & G5 V    & 2.2 \\
 217924   & G2 V    & 2.2 \\
\enddata
\tablecomments{}
\tablenotetext{a}{dwarf = luminosity classification from comparison
of absolute magnitude with canonical value rather than from the 
spectrum itself.}
\end{deluxetable}


\begin{deluxetable}{lccccccc}
\tablenum{4}
\tablewidth{0pc}
\tablecaption{Equivalent Widths  \label{tbl-EW}}
\tablehead{\colhead{Wavelength} & \colhead{EP} & \colhead{log$gf$} &
\multicolumn{5}{c}{Equivalent Width} \\
\colhead{(\AA)} & \colhead{(eV)} & \colhead{} &
\colhead{HD 100167 } & \colhead{HD 135991 }& \colhead{HD 140667 }&
\colhead{HD 158222 }& \colhead{HD 217924 }
}
\startdata
   5522.45 &      4.21&   $-$1.55 &     41.2 &            ... &           32.0 &         45.9 &        27.9 \\
   5539.28 &      3.64&   $-$2.66 &     ...  &           ...  &            8.9 &           ...  &      11.5 \\
   5543.94 &      4.22&   $-$1.14 &     ...  &           ...  &           47.9 &         62.4 &        ...\\
   5546.50 &      4.37&   $-$1.31 &     ...  &           ...  &           36.4 &           ...  &          ...\\
   5546.99 &      4.22&   $-$1.91 &     ...  &           ...  &           12.9 &           ...   &     41.1 \\
   5554.88 &      4.55&   $-$0.44 &     94.3 &           76.8 &            76.3 &       106.5 &        78.9 \\
   5560.21 &      4.43&   $-$1.19 &     46.5 &              ... &          31.5 &           ...  &          ...\\
   5576.09 &      3.43&   $-$1.00 &    113.2 &           92.5  &           97.4 &       113.6 &        98.4 \\
   5679.02 &      4.65&   $-$0.92 &    ...   &           ...  &          41.2 &           ...  &          ...\\
   5717.83 &      4.28&   $-$1.13 &    ...   &         54.4 &            37.7 &          66.0 &        59.5 \\
   5731.76 &      4.26&   $-$1.30 &    ...   &          ...  &           38.8 &            ...  &      46.8 \\
   5732.27 &      4.99&   $-$1.56 &    ...   &          ...  &            3.7  &           ...  &        ...\\
   5741.85 &      4.26&   $-$1.85 &    ...   &         21.1 &             ...  &          ...   &      27.7 \\
   5752.03 &      4.55&   $-$1.18 &    51.2  &          43.0  &           43.2 &         54.7 &        32.7 \\
   5775.08 &      4.22&   $-$1.30 &    ...   &          ...  &           46.5 &          55.8 &        52.4 \\
   5778.45 &      2.59&   $-$3.48 &    ...   &         ...  &              ...    &        ...  &      12.7 \\
   5905.67 &      4.65&   $-$0.73 &     56.2 &           48.3 &            39.8 &        61.0 &        37.0 \\
   5916.25 &      2.45&   $-$2.99 &     54.2 &           ...     &         36.8 &        54.5 &        38.8 \\
   5927.79 &      4.65&   $-$1.09 &     40.6 &           27.2  &           26.1 &        42.4 &        34.4 \\
   5929.68 &      4.55&   $-$1.41 &     38.0 &           23.0 &             ... &           ... &        ...\\
   5930.17 &      4.65&   $-$0.23 &     90.4 &           85.2  &            ...  &             ... &   82.3 \\
   5934.65 &      3.93&   $-$1.17 &     ...  &          64.7  &            ...  &             ... &    66.2 \\
   5956.69 &      0.86&   $-$4.60 &     44.1 &            ...    &         33.6 &            ... &     34.7 \\
   6003.01 &      3.88&   $-$1.12 &     67.5 &           ...     &         64.8 &        75.6 &        73.8 \\
   6027.05 &      4.08&   $-$1.09 &     59.7 &           56.2  &           46.8 &        67.5 &        42.8 \\
   6055.99 &      4.73&   $-$0.46 &    67.1  &          59.7  &            ...  &         ... &        50.8 \\
   6065.48 &      2.61&   $-$1.53 &    112.1 &           99.6   &           ...   &       ... &       100.2 \\
   6078.49 &      4.80&   $-$0.32 &     75.1 &          ...      &         58.6 &        80.7 &        58.8 \\
   6079.00 &      4.65&   $-$1.12 &     40.5 &          ...     &          29.4  &       40.5 &        27.0 \\
   6127.91 &      4.14&   $-$1.40 &      ... &            ...     &        33.0  &        ... &        36.1 \\
   6151.62 &      2.18&   $-$3.30 &     47.6 &          ...     &          31.1  &       48.9 &        34.2 \\
   6159.37 &      4.61&   $-$1.97 &     10.8 &            ...   &            ...  &       ... &          ...\\
   6165.36 &      4.14&   $-$1.47 &     38.1 &           28.6  &           25.5 &        44.7 &        31.1 \\
   6173.34 &      2.22&   $-$2.88 &      ... &           51.4  &           53.1 &         ... &        63.0 \\
   6187.99 &      3.94&   $-$1.72 &     41.3 &            ...   &           ... &        49.0 &          ...\\
   6213.43 &      2.22&   $-$2.48 &     70.8 &           58.2  &           71.1 &          ... &       68.4 \\
   6219.28 &      2.20&   $-$2.43 &     80.3 &           69.8  &           75.9 &        86.9 &        70.4 \\
   6246.32 &      3.60&   $-$0.73 &      ... &           99.1   &           ... &            ...&      94.7 \\
   6252.55 &      2.40&   $-$1.69 &    117.3 &          103.9  &           97.7 &       116.3 &          ...\\
   6265.13 &      2.17&   $-$2.55 &     78.6 &           66.9  &           65.7  &       85.6 &        75.8 \\
   6271.28 &      3.33&   $-$2.72 &      ... &            ...   &         ... &          21.1 &          ...\\
   6290.97 &      4.73&   $-$0.78 &      ... &           72.9  &            ... &          ...&          ...\\
   6293.92 &      4.83&   $-$1.72 &     38.7 &            ...   &           ... &        51.3 &          ...\\
   6322.68 &      2.59&   $-$2.43 &     ...  &           54.9    &         59.6 &          ...&        53.0 \\
   6335.33 &      2.20&   $-$2.18 &     92.5 &           76.5  &           84.4 &        95.6 &        86.2 \\
   6336.82 &      3.68&   $-$0.91 &    101.9 &           91.3  &           87.0 &       107.1 &        90.8 \\
   6344.15 &      2.43&   $-$2.92 &     ...  &            ...   &          46.7 &        64.9 &        48.8 \\
   6380.74 &      4.19&   $-$1.38 &     50.1 &           33.1   &          40.2 &          ...&          ...\\
   6393.61 &      2.43&   $-$1.57 &    120.6 &           98.8  &          104.5 &       127.6 &       107.8 \\
   6469.19 &      4.83&   $-$0.77 &     45.5 &           ...  &            40.2 &        55.2 &        56.0 \\
   6533.94 &      4.56&   $-$1.38 &     29.0 &           19.0 &             ... &        33.0 &          ...\\
   6591.31 &      4.59&   $-$2.07 &    ...   &           ... &            ...  &         14.8 &          ...\\
   6593.87 &      2.43&   $-$2.42 &     77.2 &           66.8  &           66.2 &        88.8 &        79.9 \\
   6597.56 &      4.79&   $-$1.07 &     41.8 &           33.7  &           31.5 &        44.6 &        37.4 \\
   6608.02 &      2.28&   $-$4.03 &     13.9 &            ...   &            ...  &      24.9 &          ...\\
   6677.99 &      2.69&   $-$1.35 &    126.6 &           97.1  &          106.2 &       116.1 &       102.6 \\
   6703.57 &      2.76&   $-$3.16 &     34.9 &           16.8   &          20.5 &        40.0 &          ...\\
   6710.32 &      1.49&   $-$4.88 &    ...   &           ...   &          ...  &         18.7 &          ...\\
   6715.38 &      4.61&   $-$1.64 &     24.6 &           18.8 &             ...  &       30.8 &          ...\\
   6716.22 &      4.58&   $-$1.92 &    13.5  &           8.0 &             9.4 &         15.3 &          ...\\
   6725.35 &      4.10&   $-$2.30 &     ...  &           ...  &            ...  &        16.5 &        14.1 \\
   6726.67 &      4.61&   $-$1.13 &     44.5 &           36.0  &           29.5 &        47.6 &        37.3 \\
   6750.15 &      2.42&   $-$2.62 &     68.8 &           55.5  &           53.8 &        72.5 &        63.5 \\
   6752.72 &      4.64&   $-$1.30 &      ... &           13.0   &           ... &         ... &        30.7 
\enddata
\tablecomments{Equivalent widths are given in units of m\AA.}
\end{deluxetable}


\begin{deluxetable}{lcccccc}
\tablenum{5}
\tablewidth{0pc}
\tablecaption{ Model Atmosphere Parameters and Iron Abundances \label{tab::atmosabund}}
\tablehead{
\colhead{HD} & \colhead{$T_{eff}$} & \colhead{log $g$} & \colhead{$v_{t}$}
& \colhead{[{Fe}/{H}]} & \colhead{$\sigma$} & \colhead{Number of Lines} \\
\colhead{} & \colhead{(K)} & \colhead{(cgs)} & \colhead{(km s$^{-1}$)}
& \colhead{} & \colhead{} & \colhead{}
}
\startdata
 100167   & 5900   &   4.37   &  0.5 & 0.06    & 0.14 & 41  \\
 135991   & 6050   &   4.34   &  1.0 & $-$0.17 & 0.16 & 34  \\
 140667   & 5950   &   4.52   &  1.1 & $-$0.27 & 0.11 & 44  \\
 158222   & 5700   &   4.35   &  0.6 & 0.02    & 0.18 & 40  \\
 217924   & 5900   &   4.53   &  0.5 & $-$0.12 & 0.18 & 45  \\
\enddata
\end{deluxetable}


\begin{deluxetable}{lcccc}
\tablenum{6}
\tablewidth{0pc}
\tablecaption{Iron Abundance Comparison \label{tbl-abundcomp}}
\tablehead{
\colhead{HD} &  \multicolumn{2}{c}{\citet{hetal07}} & 
 \multicolumn{2}{c}{This Work} \\
\colhead{} & \colhead{$T_{eff}$} & \colhead{[{Fe}/{H}]} & \colhead{$T_{eff}$} 
& \colhead{[Fe/H]} \\
\colhead{} & \colhead{(K)} & \colhead{} & \colhead{(K)} & \colhead{}
}
\startdata
 100167   & 5728 & $-$0.05 & 5900 & 0.06 \\
 135991   & 5929 & $-$0.20 & 6050 & $-$0.17 \\
 140667   & 5768 & $-$0.35 & 5950 & $-$0.27 \\
 158222   & 5598 & $-$0.13 & 5700 & 0.02 \\
 217924   & 5702 & $-$0.34 & 5900 & $-$0.12 \\
\enddata
\end{deluxetable}


\begin{deluxetable}{lcrrl}
\tablenum{7}
\tablewidth{0pt}
\tablecaption{Radial Velocities of HD~100167 \label{tbl-100167}}
\tablehead{ \colhead{Hel. Julian Date} & \colhead {Phase} & 
\colhead{Rad. Vel.} & \colhead{$(O-C)$} & \colhead{Source\tablenotemark{a}} \\
\colhead{($-$ 2400000)} & \colhead{} & \colhead{(km~s$^{-1}$)} &
\colhead{(km~s$^{-1}$)} & \colhead{}
}
\startdata
 52330.948  &  0.502  &  $-$28.0  &    0.3  & KPNO \\
 52707.856  &  0.724  &  $-$22.4  &    0.1  & KPNO \\
 52708.842  &  0.740  &  $-$22.0  &  $-$0.2 & KPNO \\
 52759.791  &  0.581  &  $-$26.6  &    0.2  & KPNO \\
 53120.734  &  0.539  &  $-$27.4  &    0.3  & KPNO \\
 53171.678  &  0.380  &  $-$29.6  &    0.1  & KPNO \\
 53491.705  &  0.663  &  $-$24.5  &    0.2  & KPNO \\
 53532.642  &  0.338  &  $-$29.9  &    0.1  & KPNO \\
 53536.648  &  0.404  &  $-$29.6  &  $-$0.1 & KPNO \\
 53850.791  &  0.590  &  $-$26.5  &    0.1  & KPNO \\
 54130.025  &  0.199  &  $-$29.1  &    0.1  & Fair \\
\enddata
\tablecomments{}
\tablenotetext{a}{KPNO = Kitt Peak National Observatory, 
Fair = Fairborn Observatory \\
(This table is available in its entirety in machine-readable and Virtual
Observatory (VO) forms in the online journal. A portion is shown here
for guidance regarding its form and content.)
}
\end{deluxetable}


\begin{deluxetable}{lcrrl}
\tablenum{8}
\tablewidth{0pt}
\tablecaption{Radial Velocities of HD~135991 \label{tbl-135991}}
\tablehead{ \colhead{Hel. Julian Date} & \colhead {Phase} & 
\colhead{Rad. Vel.} & \colhead{$(O-C)$} & \colhead{Source\tablenotemark{a}} \\
\colhead{($-$ 2400000)} & \colhead {} & \colhead{(km~s$^{-1}$)} &
\colhead{(km~s$^{-1}$)} & \colhead{}
}
\startdata
 52392.889  & 0.619  & $-$21.1  &   0.1  & KPNO \\
 52757.857  & 0.036  & $-$36.4  & $-$0.2 & KPNO \\
 52759.839  & 0.049  & $-$35.7  &   0.0  & KPNO \\
 53121.900  & 0.447  & $-$23.4  & $-$0.3 & KPNO \\
 53168.803  & 0.758  & $-$20.2  &   0.0  & KPNO \\
 53492.873  & 0.904  & $-$21.8  & $-$0.2 & KPNO \\
 53532.786  & 0.168  & $-$29.5  & $-$0.3 & KPNO \\
 53535.824  & 0.188  & $-$28.7  & $-$0.2 & KPNO \\
 53851.874  & 0.281  & $-$26.0  & $-$0.1 & KPNO \\
 54003.590  & 0.286  & $-$25.8  &   0.0  & KPNO \\
 54221.797  & 0.731  & $-$20.4  & $-$0.1 & KPNO \\
 54246.744  & 0.896  & $-$21.4  & $-$0.1 & Fair \\
\enddata
\tablecomments{}
\tablenotetext{a}{KPNO = Kitt Peak National Observatory,
Fair = Fairborn Observatory \\
(This table is available in its entirety in machine-readable and Virtual
Observatory (VO) forms in the online journal. A portion is shown here
for guidance regarding its form and content.)
}
\end{deluxetable}


\begin{deluxetable}{lcrrl}
\tablenum{9}
\tablewidth{0pt}
\tablecaption{Radial Velocities of HD~140667 \label{tbl-140667}}
\tablehead{ \colhead{Hel. Julian Date} & \colhead {Phase} & 
\colhead{Rad. Vel.} & \colhead{$(O-C)$} & \colhead{Source\tablenotemark{a}} \\
\colhead{($-$ 2400000)} & \colhead{} & \colhead{(km~s$^{-1}$)} &
\colhead{(km~s$^{-1}$)} & \colhead{}
}
\startdata
 52706.010  & 0.616  & $-$14.8  & $-$0.2 & KPNO \\
 52755.944  & 0.667  & $-$15.8  & $-$0.2 & KPNO \\
 53119.928  & 0.039  & $-$21.1  & $-$0.1 & KPNO \\
 53169.792  & 0.090  & $-$19.4  & $-$0.2 & KPNO \\
 53491.880  & 0.419  & $-$13.1  & $-$0.2 & KPNO \\
 53532.809  & 0.461  & $-$13.1  & $-$0.1 & KPNO \\
 53851.895  & 0.787  & $-$18.6  &   0.0  & KPNO \\
 53904.689  & 0.841  & $-$20.2  &   0.0  & Fair \\
 53927.726  & 0.865  & $-$20.9  & $-$0.1 & Fair \\
 53998.650  & 0.937  & $-$22.5  & $-$0.3 & Fair \\
\enddata
\tablecomments{}
\tablenotetext{a}{KPNO = Kitt Peak National Observatory,
Fair = Fairborn Observatory \\
(This table is available in its entirety in machine-readable and Virtual
Observatory (VO) forms in the online journal. A portion is shown here
for guidance regarding its form and content.)
}
\end{deluxetable}


\begin{deluxetable}{lcrrl}
\tablenum{10}
\tablewidth{0pt}
\tablecaption{Radial Velocities of HD~158222 \label{tbl-158222}}
\tablehead{ \colhead{Hel. Julian Date} & \colhead {Phase} & 
\colhead{Rad. Vel.} & \colhead{$(O-C)$} & \colhead{Source\tablenotemark{a}} \\
\colhead{($-$ 2400000)} & \colhead{} & \colhead{(km~s$^{-1}$)} &
\colhead{(km~s$^{-1}$)} & \colhead{}
}
\startdata
 52538.647  & 0.409  & $-$29.2  &   0.3  & KPNO \\
 52758.966  & 0.478  & $-$27.6  & $-$0.1 & KPNO \\
 53171.800  & 0.481  & $-$27.6  & $-$0.2 & KPNO \\
 53492.910  & 0.039  & $-$29.5  & $-$0.2 & KPNO \\
 54003.630  & 0.517  & $-$26.1  &   0.1  & KPNO \\
 54222.889  & 0.581  & $-$24.3  & $-$0.1 & KPNO \\
 54238.685  & 0.658  & $-$21.6  &   0.0 & Fair \\
 54240.686  & 0.668  & $-$21.3  &   0.0 & Fair \\
 54241.686  & 0.673  & $-$21.1  &   0.0 & Fair \\
 54251.963  & 0.722  & $-$19.1  &   0.2 & Fair \\
\enddata
\tablecomments{}
\tablenotetext{a}{KPNO = Kitt Peak National Observatory,
Fair = Fairborn Observatory \\
(This table is available in its entirety in machine-readable and Virtual
Observatory (VO) forms in the online journal. A portion is shown here
for guidance regarding its form and content.)
}
\end{deluxetable}


\begin{deluxetable}{lcrrl}
\tablenum{11}
\tablewidth{0pt}
\tablecaption{Radial Velocities of HD~217924 \label{tbl-217924}}
\tablehead{ \colhead{Hel. Julian Date} & \colhead {Phase} & 
\colhead{Rad. Vel.} & \colhead{$(O-C)$} & \colhead{Source\tablenotemark{a}} \\
\colhead{($-$ 2400000)} & \colhead{} & \colhead{(km~s$^{-1}$)} &
\colhead{(km~s$^{-1}$)} & \colhead{} 
}
\startdata
 52540.820 &  0.901  & 20.1  & 0.0    & KPNO \\ 
 52902.791 &  0.052  & 12.7  & $-$0.1 & KPNO \\ 
 52941.726 &  0.068  & 13.4  & 0.4    & KPNO \\
 53168.985 &  0.162  & 13.5  & $-$0.4 & KPNO \\ 
 53273.772 &  0.206  & 14.1  & $-$0.2 & KPNO \\  
 53531.966 &  0.314  & 14.7  & $-$0.3 & KPNO \\  
 53639.786 &  0.358  & 15.0  & $-$0.2 & KPNO \\    
 54003.751 &  0.510  & 16.1  &   0.0  & KPNO \\   
 54006.771 &  0.511  & 16.1  &  0.0   & Fair \\ 
 54050.649 &  0.529  & 16.2  &  0.0   & Fair \\
\enddata
\tablecomments{}
\tablenotetext{a}{KPNO = Kitt Peak National Observatory,
Fair = Fairborn Observatory \\
(This table is available in its entirety in machine-readable and Virtual
Observatory (VO) forms in the online journal. A portion is shown here
for guidance regarding its form and content.)
}
\end{deluxetable}


\begin{figure} 
\figurenum{1} 
\epsscale{0.8} 
\plotone{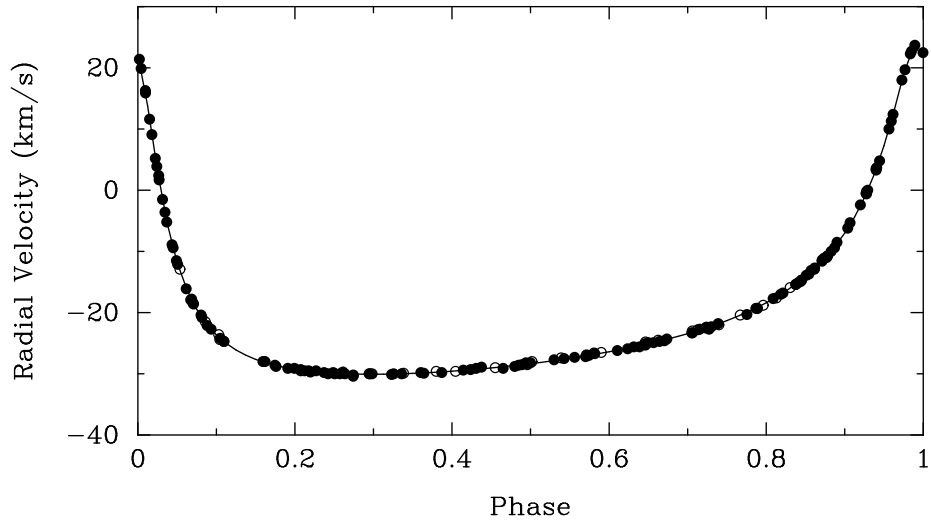} 
\figcaption{Radial velocities of HD 100167 compared with the computed
velocity curve.  Dots = Fairborn Observatory, open circles = KPNO.
Zero phase is a time of periastron passage. 
\label{fig:100167orb}
}
\end{figure}


\begin{figure}
\figurenum{2}
\epsscale{0.8}
\plotone{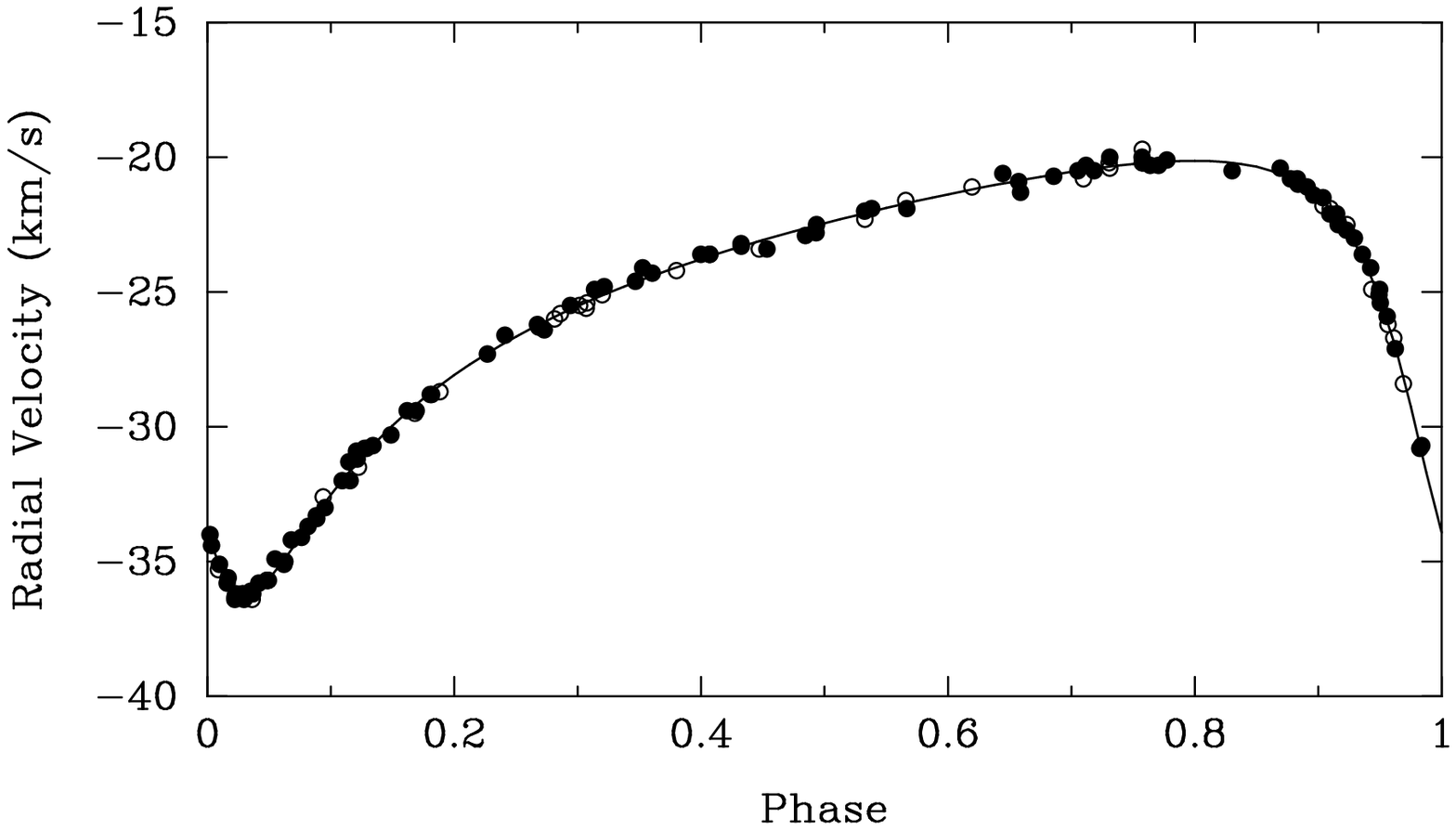}
\figcaption{Radial velocities of HD 135991 compared with the computed
velocity curve.  Dots = Fairborn Observatory, open circles = KPNO.
Zero phase is a time of periastron passage.
\label{fig:135991orb}
}
\end{figure}


\begin{figure}
\figurenum{3}
\epsscale{0.8}
\plotone{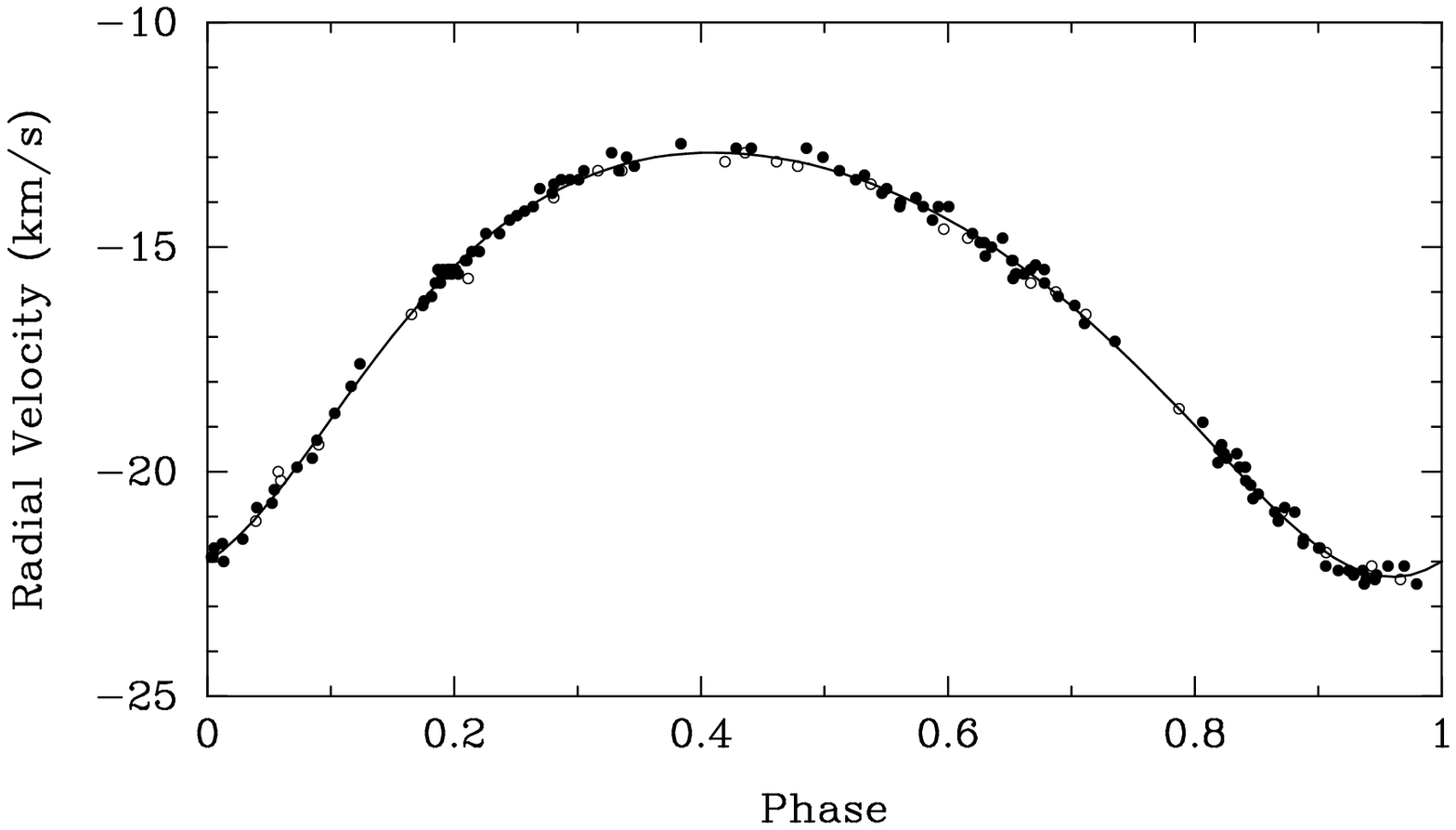}
\figcaption{Radial velocities of HD 140667 compared with the computed
velocity curve.  Dots = Fairborn Observatory, open circles = KPNO.
Zero phase is a time of periastron passage.
\label{fig:140667orb}
}
\end{figure}


\begin{figure}
\figurenum{4}
\epsscale{0.8}
\plotone{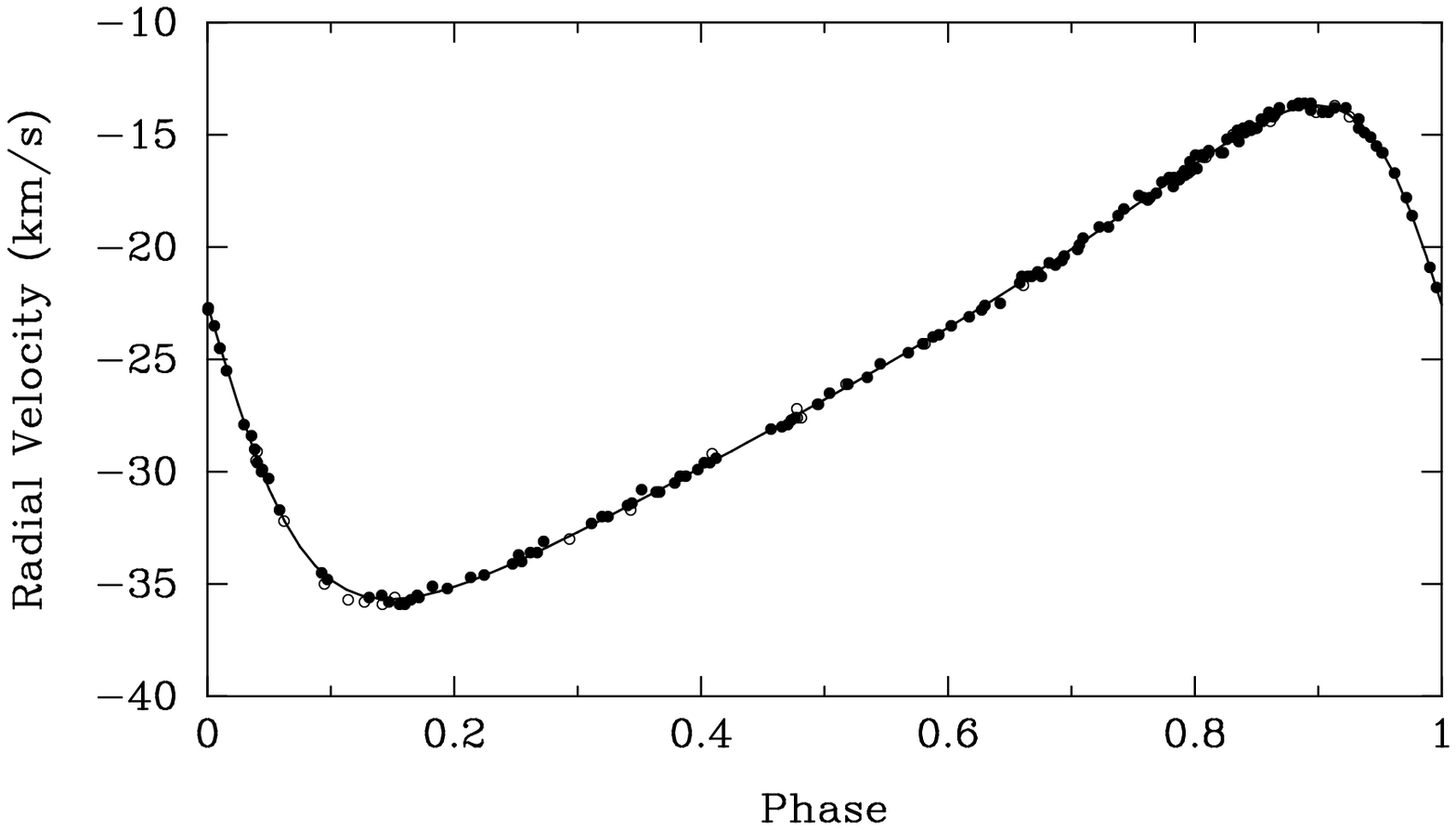}
\figcaption{Radial velocities of HD 158222 compared with the computed
velocity curve.  Dots = Fairborn Observatory, open circles = KPNO.
Zero phase is a time of periastron passage.
\label{fig:158222orb}
}
\end{figure}


\begin{figure}
\figurenum{5}
\epsscale{0.8}
\plotone{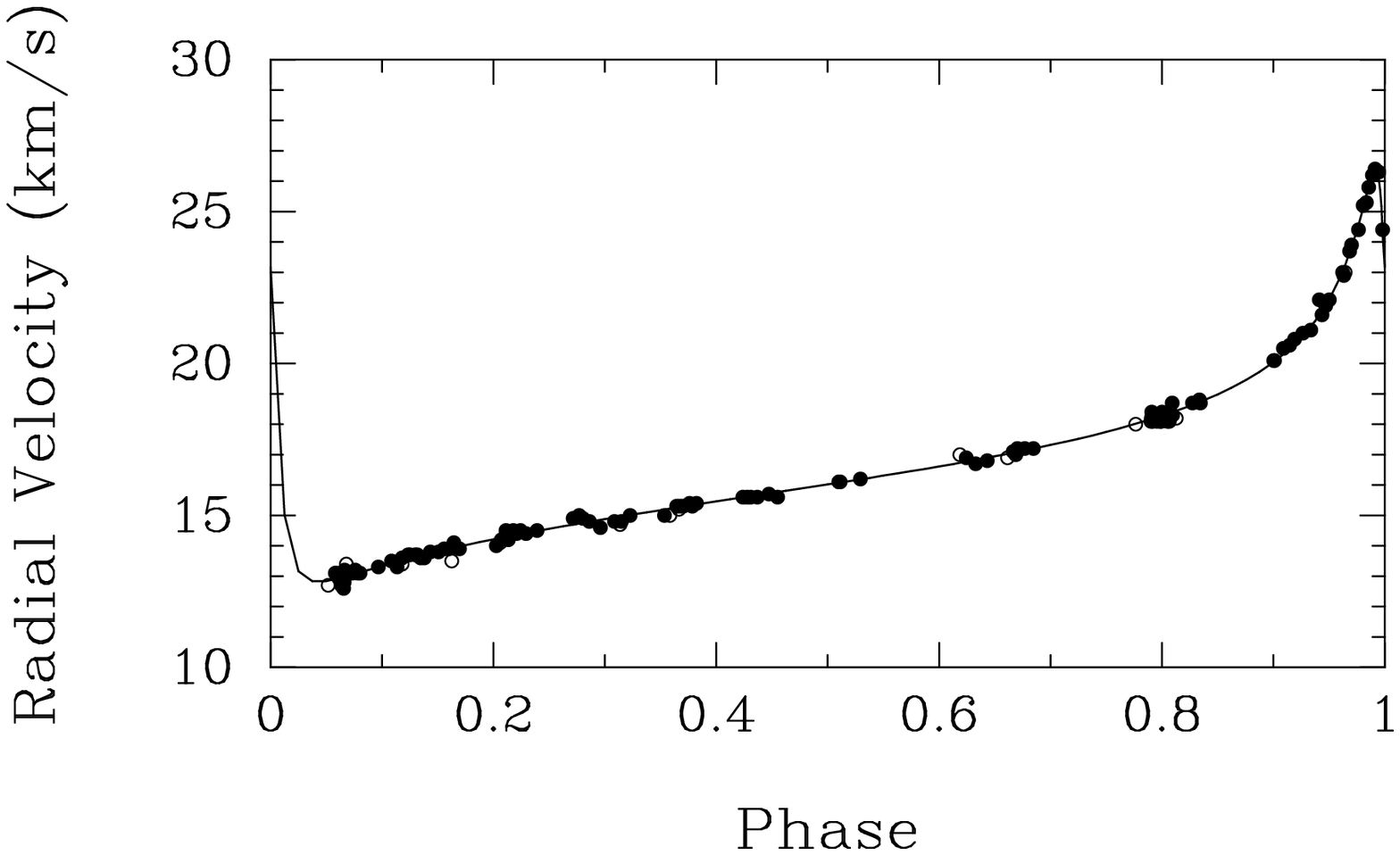}
\figcaption{Radial velocities of HD 217924 compared with the computed
velocity curve.  Dots = Fairborn Observatory, open circles = KPNO.
Zero phase is a time of periastron passage.
\label{fig:217924orb}
}
\end{figure}

\end{document}